\documentclass[runningheads]{llncs}
\usepackage{graphicx}
\usepackage{todonotes}
\usepackage{amsmath}
\usepackage{braket}
\usepackage{tikz}
\usepackage{hyperref}
\newcommand\copyrighttext{%
  \footnotesize This is a pre-print of an article to be published in \emph{Feld S., Linnhoff-Popien C. (eds)}, Quantum Technology and Optimization Problems (QTOP 2019), Lecture Notes in Computer Science, vol 11413, Springer, 2019. The final authenticated version is available online at DOI:  \href{http://doi.org/10.1007/978-3-030-14082-3}{http://doi.org/10.1007/978-3-030-14082-3}}
\newcommand\copyrightnotice{%
\begin{tikzpicture}[remember picture,overlay]
\node[anchor=south,yshift=10pt] at (current page.south) {\fbox{\parbox{\textwidth}{\copyrighttext}}};
\end{tikzpicture}%
}

\begin{document}
\title{Assessing Solution Quality of 3SAT\\on a Quantum Annealing Platform}
%
%
\author{Thomas Gabor\inst{1} \and
Sebastian Zielinski\inst{1} \and
Sebastian Feld\inst{1} \and
Christoph Roch\inst{1} \and
Christian Seidel\inst{2} \and
Florian Neukart\inst{3} \and
Isabella Galter\inst{2} \and\\
Wolfgang Mauerer\inst{4} \and
Claudia Linnhoff-Popien\inst{1}}
\authorrunning{Gabor et al.}
%
\institute{LMU Munich\and
Volkswagen Data:Lab \and
Volkswagen Group of America\and
OTH Regensburg/Siemens Corporate Research}
\maketitle              

\copyrightnotice{}
\begin{abstract}
When solving propositional logic satisfiability (specifically 3SAT) using quantum annealing, we analyze the effect the difficulty of different instances of the problem has on the quality of the answer returned by the quantum annealer. A high-quality response from the annealer in this case is defined by a high percentage of correct solutions among the returned answers. We show that the phase transition regarding the computational complexity of the problem, which is well-known to occur for 3SAT on classical machines (where it causes a detrimental increase in runtime), persists in some form (but possibly to a lesser extent) for quantum annealing.

\end{abstract}
\section{Introduction}


Quantum computers are an emerging technology and still subject to frequent new developments. Eventually, the utilization of intricate physical phenomena like superposition and entanglement is conjectured to provide an advantage in computational power over purely classical computers. As of now, however, the first practical breakthrough application for quantum computers is still sought for. But new results on the behavior of quantum programs in comparison to their classical counterparts are reported on a daily basis.

Research in that area has cast an eye on the complexity class NP: It contains problems that are traditionally (and at the current state of knowledge regarding the P vs.\ NP problem) conjectured to produce instances too hard for classical computers to solve exactly and deterministically within practical time constraints. Still, problem instances of NP are also easy enough that they can be executed efficiently on a (hypothetical) non-deterministic computer.

The notion of computational complexity is based on classical
computation in the sense of using classical mechanics to describe and
perform automated computations. In particular, it is known that in this
model of computation, simulating quantum mechanical systems is hard. However, nature itself routinely ``executes'' quantum mechanics, leading to
speculations~\cite{feynman1981simulating} that quantum mechanics may be used to leverage 
 greater computational power than systems adhering to the rules of classical physics
can provide.

Quantum computing describes technology exploiting the behavior of quantum mechanics to build computers that are (hopefully) more powerful than current classical machines. Instead of classical bits $b \in \{0, 1\}$ they use qubits $q = \alpha\ket{0} + \beta\ket{1}$ where $\alpha, \beta, |\alpha|^2 + |\beta|^2 = 1,$ are probability amplitudes for the basis states $\ket{0}, \ket{1}$. Essentially, a qubit can be in both states $0$ and $1$ at once, each with a specific probability. This phenomenon is called superposition, but it collapses when the actual value of qubit is measured, returning either $0$ or $1$ with said specific probability and fixing that randomly acquired result as the future state of the qubit. Entanglement describes the effect that multiple quits can be in superpositions that are affected by each other, meaning that the measurement of one qubit can change the assigned probability amplitudes of another qubit in superposition. The combination of these phenomena allows qubits to concisely represent complex data and lend themselves to efficient computation operations.

In this work, we focus on the concrete technological platform of quantum annealing that is (unlike the generalized concept of quantum computing) not capable of executing general quantum mechanical computations, but is within current technological feasibility, and available to researchers outside the field of quantum hardware. The mechanism specializes in solving optimization problems, and can (as a trade-off) work larger amounts of qubits in a useful way than quantum mechanically complete platforms.

In this paper, we evaluate the performance of quantum annealing (or more specifically, a D-Wave 2000Q machine) on the canonical problem of the class NP, propositional logic satisfiability for 3-literal clauses (3SAT)~\cite{cook1971complexity}. As we note that there is still a remarkable gap between 3SAT instances that can be put on a current D-Wave chip and 3SAT instances that even remotely pose a challenge to classical solvers, there is little sense in comparing the quantum annealing method to classical algorithms in this case (and at this early point in time for the development of quantum hardware). Instead, we are interested in the scaling behavior with respect to problem difficulty. Or more precisely: We analyze if and to what extent quantum annealing's performance suffers under hard problem instances (like classical algorithms do).

We present a quick run-down of 3SAT and the phenomenon of phase transitions in Section~\ref{sec:preliminaries} and continue to discuss further related work in Section~\ref{sec:related-work}. In Section~\ref{sec:exp-setup} we describe our experimental setup and then present the corresponding results in Section~\ref{sec:evaluation}. We conclude with Section~\ref{sec:conclusion}.

\section{Preliminaries}
\label{sec:preliminaries}

Propositional logic satisfiability (SAT) is the problem of telling if a given formula in propositional logic is satisfiable, i.e., if there is a assignment to all involved Boolean variables that causes the whole formula to reduce to the logical value $\textit{True}$. As such, the problem occurs at every application involved complex constraints or reasoning, like (software) product lines, the tracing of software dependencies or formal methods.

It can be trivially shown that (when introducing a linear amount of new variables) all SAT problems can be reduced to a specific type of SAT problem called 3SAT, where the input propositional logic formula has to be in conjunctive normal form with all of the disjunctions containing exactly three literals.

For example, the formula $\Psi = ( x_{1} \vee x_{2} \vee x_{3}) \wedge ( \lnot x_{1} \vee x_{2} \vee x_{3})$ is in 3SAT form and is satisfiable because the assignment $(x_1 \mapsto \textit{True}, x_2 \mapsto \textit{True}, x_2 \mapsto \textit{True})$ causes the formula to reduce to $\textit{True}$. The formula $\Phi = ( x_{1} \vee x_{1} \vee x_{1}) \wedge ( \lnot x_{1} \vee \lnot x_{1}  \vee \lnot x_{1} )$ is also in 3SAT form but is not satisfiable.

\subsubsection{Definition (3SAT)}
A 3SAT instance with $m$ clauses and $n$ variables is given as a list of clauses $(c_k)_{0 \leq k \leq m-1}$ of the form $c_k = (l_{3k} \lor l_{3k+1} \lor l_{3k+2})$ and a list of variables $(v_j)_{0 \leq j \leq n-1}$ so that $l_i$ is a literal of the form $l_i \in \bigcup_{0 \leq j \leq n-1} \{v_j, \lnot v_j\}$. A given 3SAT instance is \emph{satisfiable} iff there exists a variable assignment $(v_j \mapsto b_j)_{0 \leq j \leq n-1}$ with $b_j \in \{\textit{True}, \textit{False}\}$ so that $\bigwedge_{0 \leq k \leq m-1} c_k$ reduces to $\textit{True}$ when interpreting all logical operators as is common. The problem of deciding whether a given 3SAT instance is satisfiable is called 3SAT.\\

3SAT is of special importance to complexity theory as it was the first problem which was shown to be NP-complete \cite{cook1971complexity}. This means that every problem in NP can be reduced to 3SAT in polynomial time. It follows that any means to solve 3SAT efficiently would thus give rise to efficient solutions for any problem in NP like graph coloring, travelling salesman or bin packing.

%
%
%

Despite the fact that for NP-complete problems in general no algorithm is known that can solve all problem instances of a problem efficiently (i.e., in polynomial time), it is within the scope of knowledge that ``average'' problem instances of many NP-complete problems, including 3SAT, are easy to solve~\cite{cheeseman1991really}. In Ref.~\cite{monasson1999determining} this characteristic is described with a phase transition. The boundary of the phase transition divides the problem space into two regions. In one region, a solution can be found relatively easily, because the solution density for these problems is high, whereas in the other region, it is very unlikely that problems can contain a correct solution at all. Problems that are very difficult to solve are located directly at this phase boundary~\cite{cheeseman1991really}.

\begin{figure}[t]
	\centering
	\input{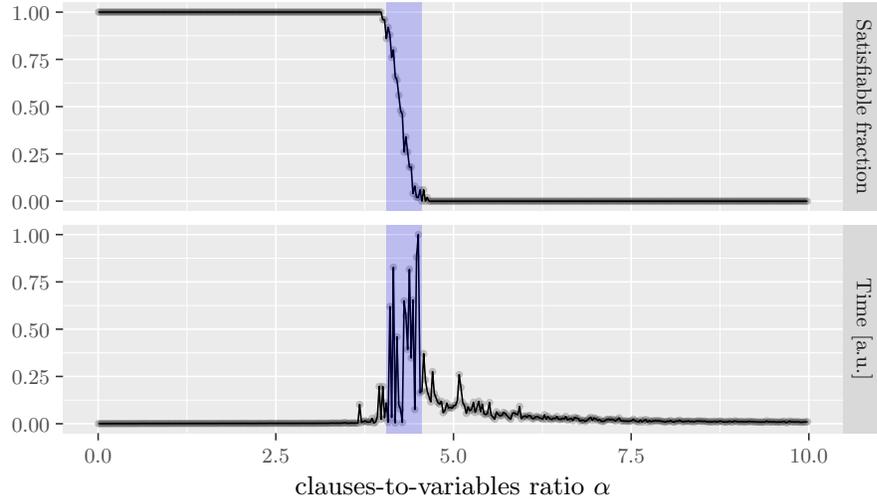}
	\caption{Phase transition of SAT. The bottom plot shows the computational time required to determine satisfiability of randomly chosen 3SAT instances with specific a clauses-to-variables ratio $\alpha$ on a standard solver. The area around the critical point $\alpha_{c} \approx 4.267$ is shaded in blue.\newline
          The upper portion shows the probability that instances with a particular ratio $\alpha$ are solvable. In the region around the critical point, it is hard to determine whether a problem instance can be fulfilled with a concrete allocation or not.} \label{fig:crit_sat} \end{figure}

It can be observed that, with randomly generated 3SAT instances, the probability of finding a correct solution decreases abruptly when the ratio of clauses to variables $\alpha = m/n$ exceeds a critical value of $\alpha_{c}$ \cite{monasson1996entropy}. According to \cite{mezard2002random} this critical point is $\alpha_{c} \approx $ 4.267 for randomly generated 3SAT instances. In the surrounding area of the critical point, finding a solution (i.e., deciding if the instance is satisfiable) is algorithmically complex. Figure~\ref{fig:crit_sat} illustrates this phenomenon.

To assess the solution quality of randomly generated 3SAT instances we generate instances in every complexity region. The results are discussed in Section~\ref{sec:evaluation}.

\section{Related Work}
\label{sec:related-work}

It is one of the cornerstones of complexity theory that solving NP-complete or even NP-hard decision problems is strongly believed to be not efficiently possible~\cite{cook1971complexity,murty1987some}. Any NP-complete problem can also be cast as an optimization problem, which allows for employing well-known optimization algorithms to find approximate solutions---typical methods include tabu search \cite{glover2013tabu,gendreau1994tabu} and simulated annealing \cite{kirkpatrick1983optimization,chen1995chaotic}. Countless other efficient approximation methods, together with an elaborate taxonomy on approximation quality (how much does a given solution differ from a known global optimum?) and computational effort (how many time steps are required until an approximate solution that satisfies given quality goals is available?), have been devised \cite{Ausiello1999}.

Some problem (knapsack, e.g.) exhibit favorable properties when cast as an optimization problem. The latter is a member of the complexity class FPTAS (fully polynomial-time approximation scheme), which means that a solution with distance \(1+\epsilon\) (of course, $\epsilon > 0$) from an optimal solution can be determined in polynomial time in both, input size \(n\) and inverse approximation quality \(1/\epsilon\)~\cite{chen1995chaotic}.

An intriguing connection that has received substantial attraction exists between (computational) NP-complete problems and the (physical) concept of phase transitions, as detailed in Section~\ref{sec:preliminaries}. First investigations of the phenomenon have been performed by Kirkpatrick et al.~\cite{kirkpatrick1994critical}; Monasson et al.\ first suggested a connection between the type of phase transition and the associated computational costs of a problem~\cite{monasson1999determining}. From the abundant amount of more recent investigations, we would like to highlight the proof by Ding et al.~\cite{ding2015proof} that establishes a threshold value for the phase transition. Our work benefits from the above insights by selecting the ``most interesting'', i.e., computationally hardest, scenarios as investigation target.

The idea of obtaining solutions for NPO (NP optimization) problems by finding the energy ground state (or states) of a quantum mechanical system was used, for instance, by Apolloni et al.~\cite{apolloni1989quantum,apolloni1988numerical} to solve combinatorial optimization problems. The general idea of quantum annealing has been independently re-discovered multiple times~\cite{albash2016adiabatic,finnila1994quantum,amara1993global,kadowaki1998quantum}.

Quantum annealing techniques are usually applied to solving NP-complete or NP-hard decision problems, or optimization problems from class NPO. Lucas~\cite{lucas2014ising} reviews how to formulate a set of key NP problems in the language of adiabatic quantum computing respectively quadratic unconstrained binary optimization (QUBO). In particular, problems of the types ``travelling salesman'' or ``binary satisfiability'' that are expected to have a major impact on practical computational applications if they can be solved advantageously on quantum annealers have undergone a considerable amount of research~\cite{heim2017designing,warren2017small,moylett2017quantum,strand2017zzz,benjamin2017measurement,neukart2017traffic}. Further effort has been made on combining classical and quantum methods on these problems~\cite{feld2018hybrid}.

Comparing the computational capabilities of classical and quantum computers is an intriguing and complex task, since the deployed resources are typically very dissimilar. For instance, the amount of instructions required to execute a particular algorithm is one of the main measures of efficiency or practicability on a classical machine, whereas the notion of a discrete computational ``step'' is hard to define on a quantum annealing device. Interest in quantum computing has also spawned definitions of new complexity classes (e.g., \cite{klauck2017complexity,morimae2017merlinization}), whose relations to traditional complexity classes have been and are still subject to ongoing research~\cite{bernstein1997quantum,marriott2005quantum}.


These questions hold regardless of any specific physical or conceptual implementation of quantum computing since their overall computational capabilities are known to be largely interchangeable; for instance, McGeoch~\cite{mcgeoch2014adiabatic} discusses the
equivalence of gate-based and adiabatic quantum computing. Consequently,
our work focuses not on comparing quantum and classical aspects of solving particular problems, but concentrates on understanding peculiarities
of solving one particular problem (3SAT, in our case) in-depth.

Formulating 3SAT problems on a quantum annealing hardware has been previously considered~\cite{choi2011different,choi2010adiabatic,farhi2000quantum}, and we rely on the encoding techniques presented there. Van~\cite{van2001powerful} and Farhi~\cite{farhi2009quantum} have worked on analyzing the complexity of solving general 3SAT problems. Hsu et al. have considered the complexity-wise easier variation 2SAT as a benchmarking problem to compare various parameter configurations of their quantum annealer~\cite{hsu2018quantum}.

\section{Experimental Setup}
\label{sec:exp-setup}

Quantum annealing is an optimization process that can be implemented in hardware. It is built upon the adiabatic theorem that provides conditions under which an initial ground-state configuration of a system evolves to the ground state of another configuration that minimizes a specific user-defined energy function~\cite{mcgeoch2014adiabatic}. As in the real world the required conditions for the theorem can only be approximated, the results of quantum annealing are usually not deterministically optimal but show a probabilistic distribution, ideally covering the desired optimal value as well.

D-Wave's quantum annealer is the first commercial machine to implement quantum annealing. Its interface is built on two equivalent mathematical models for optimization problems called Ising and QUBO, the latter of which will be used for the work of this paper. Quadratic Unconstrained Binary Optimization (QUBO) problems can be formulated as a quadratic matrix $Q_{ij}$. Quantum annealing then searches for a vector $x \in \{0,1\}^n$ so that $\sum_i \sum_{j < i} Q_{ij} x_i x_j + \sum_i Q_i x_i$ is minimal. The promise of quantum annealing is that---using quantum effects---specialized hardware architectures are able to solve these optimization problems much faster than classical computers in the future.

The main goal of this paper is to analyze the inherently probabilistic distribution of return values generated by quantum annealing when trying to solve hard optimization problems. We choose to demonstrate such an analysis on 3SAT because it is the canonical problem of the class NP, which is a prime target for research on performance improvements via quantum technology with respect to classical computers~\cite{mcgeoch2013experimental,lucas2014ising}.

\subsection{Defining 3SAT as a QUBO}
3SAT is usually not formulated as an optimization problem (see Section~\ref{sec:preliminaries}), or defined by an equivalent QUBO problem,
as is required by the annealer. Thus, we require a (polynomial-time) translation of any 3SAT instance into a QUBO so that the solutions generated by the quantum annealer can be translated back to solutions of the initial 3SAT instance.

Following~\cite{choi2010adiabatic,choi2011different}, we translate 3SAT into the Weighted Maximum Independent Set (WMIS) problem and then translate the WMIS instance into a QUBO (we find it convenient to specify the polynomial coefficients in matrix form). We omit the details of this process and instead refer to \emph{op.~cit.} and Lucas~\cite{lucas2014ising}. However, we shall briefly discuss the implications of the translation process.

A 3SAT instance, that is, a formula with $m$ clauses for $n$ variables,
requires a QUBO matrix of size $3m \times 3m$ with the solution vector $x \in \{0,1\}^{3m}$. The solution can be thought of as using a qubit for each literal in the initial formula and thus consisting of a triplet of qubits for each 3SAT clause. This usually means that we have much more qubits than variables in the formula. Nonetheless, a QUBO solution is mapped to a value assignment for the variables in the 3SAT formula. Thus, when running successfully, the quantum annealer will output a satisfying assignment for a given 3SAT formula. We can check if the assignment really is correct (i.e., each variable has a value assigned and the whole formula reduces to $\textit{True}$) using few instructions of classical computation. Obviously, if among several experimental runs the quantum annealer does return just one correct assignment, the corresponding 3SAT formula is satisfiable. If the quantum annealer only returns incorrect assignments, we will regard the formula as unsatisfiable (although the prove of that is only probabilistic).

There are some aspects to note about how the QUBO solution vectors are mapped to variable assignments. Given a QUBO solution vector $(x_i)_{0 \leq i \leq 3m-1}$ for a 3SAT formula with literals $(l_i)_{0 \leq i \leq 3m-1}$, a variable $v$ is assigned the value $\textit{True}$ if it occurs in a literal $l_i = v$ and $x_i = 1$. Likewise, a variable $v$ is assigned the value $\textit{False}$ if it occurs in a literal $l_i = \lnot v$ and $x_i = 1$. It is important to note that $x_i = 0$ has \emph{no implication} on the value of the variable in $l_i$.

Intuitively, we can interpret $x_i = 1$ to mean ``use the value of $l_i$ to prove the satisfaction of clause $c_{(i \mod 3)}$''. From our QUBO optimization, we expect to find one (and only one) suitable $l_i$ for every clause in the 3SAT formula.\footnote{This intuition matches the concept of constructivism in logic and mathematics. We are not only looking for the correct answer, but are looking for a correct and complete proof of an answer, giving us a single witness for each part of the formula.}

This is important as it opens up a wide range of different QUBO solutions which may just encode the exact same variable assignment at the 3SAT level. However, it also means that seemingly suboptimal QUBO solutions may encode correct 3SAT assignments. For example, consider the (a little redundant) 3SAT formula $(v_0 \lor v_1 \lor v_2) \land (v_0 \lor v_1 \lor v_2)$: The QUBO solution $x = 100001$ would imply the assignment of $v_0 = \textit{True}$ and $v_2 = \textit{True}$, which indeed is theoretically sufficient to prove the formula satisfiable. The exact same assignment would be implied by $x = 001100$. However, note that none of these imply a full assignment of every variable in the 3SAT instance since none say anything about the value of $v_1$. Still, we can trivially set $v_1$ to any arbitrary value and end up with a correct assignment. Also note that while the QUBO is built in such a way to opt for one single value $1$ per triplet in the bit string, even bitstrings violating this property can encode correct solution. In our example, the suboptimal QUBO solution $x=100000$ still encodes all necessary information to prove satisfiability.

\subsection{Evaluating Postprocessing}

As can be seen from the last example, postprocessing is an integral part of solving problems with quantum annealing. As discussed earlier in this section, we consider a QUBO solution correct, if it not only matches the expected structure for minimizing the QUBO energy function, but instead iff it directly implies a correct assignment in the definition given above. Thus, while the expected structure for QUBO optimizes $x$ so that the amount bits $x_i$ assigned $1$ equals the amount of clauses $m$, we also consider less full answers correct.

On top of that, there are solutions that cannot be mapped to an assignment immediately, but still with almost no effort. We want to regard these as well and implemented a postprocessing step we call \emph{logical postprocessing}. It is applied whenever none of the qubits corresponding to a single clause $c_k$ are set to $1$ by the quantum annealer and the respective QUBO solution is not already correct. In that case, we iterate through all literals $l_i$ in that clause $c_k$ and check if we could set $x_i = 1$ without contradicting any other assignment made within $x$. If we find such an $l_i$, we set $x_i = 1$ and return the altered bitstring $x$.

The software platform provided by D-Wave to use the quantum annealer already offers integrated postprocessing methods as well, which we will also empirically show to be more powerful than logical postprocessing in the following Section~\ref{sec:evaluation}. Again, for greater detail we refer to the D-Wave documentation on that matter~\cite{dwavepostprocessing}. At a glance, the employed postprocessing method splits the QUBO matrix into several subproblems, tries to optimize these locally, and then integrates that local solution into the complete solution if it yields an improvement. We call this method \emph{D-Wave postprocessing}.

To evaluate the solution quality regarding 3SAT, we employ both methods. The goal is to assess the expected quality on a 3SAT-to-3SAT level, that is, we measure how well we can solve the given 3SAT instance and regard the translation to and from QUBO as a mere technical problem that is not of
interest for this paper.

\section{Evaluation}
\label{sec:evaluation}

To assess the solution quality of 3SAT on a quantum annealing platform, using the previously discussed method of encoding 3SAT problems, we ran several experiments on a D-Wave 2000Q system. Using ToughSAT\footnote{\url{https://toughsat.appspot.com/}} we generated 3SAT instances of various difficulty (i.e., with various values for $\alpha$). However, as discussed in Section~\ref{sec:preliminaries}, for $|\alpha - 4.2| \gg 0$ problem instances become very easy to solve. We observed that effect on the quantum annealer as well, since all of these instances were easily solved on the D-Wave machine. Thus, for the remainder of this section, we focus on hard instances (approximated by $\alpha = 4.2$) to assess solution quality in the interesting problem domain.

Experiments have shown that using the standard embedding tools delivered with the D-Wave platform, we can only reliably find a working embedding on the D-Wave 2000Q chip for 3SAT instances with at most $42$ clauses~\cite{adams1995hitchhikers}. To maintain $\alpha \approx 4.2$, the generated 3SAT instances contain $10$ different variables. We only assess solution quality for 3SAT instances that are satisfiable, but do not
provide this information to the solver.

\begin{figure}[t]
\centering
\includegraphics[width=.65\textwidth]{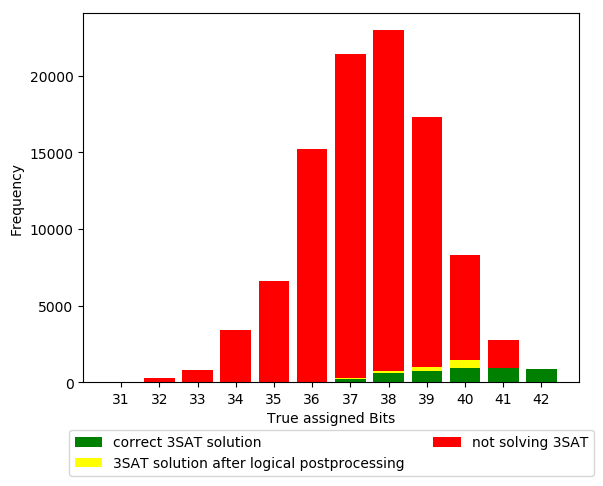}
\caption{Distribution of correct (green) and incorrect (red) answers returned by the quantum annealer \emph{without D-WAVE postprocessing}. Answers that can trivially be transformed into valid answers using logical postprocessing are marked in yellow. The plot shows 100,000 answers in total for 100 different hard 3SAT instances ($\alpha \approx 4.2$).}
\label{fig:distr-no-pp}
\end{figure}

Figure~\ref{fig:distr-no-pp} shows the result distribution of these runs on the D-Wave machine. On the x-axis, we sorted the returned results according to the bits that have been assigned the value $1$ or $\textit{True}$. As discussed in Section~\ref{sec:exp-setup} the optimal solution is supposed to set one bit for each clause, i.e., is supposed to contain 42 bits set to $\textit{True}$. However, as there are only 10 different variables, there theoretically exist answers that only set 10 bits but that still map to a complete and valid solution for the given 3SAT instance. From Figure~\ref{fig:distr-no-pp} we can see that some of these solution are found for bitcounts starting from 37 through 41. Interestingly, the complete range of answers gathered seems to follow a distribution centered around $37$ or $38$ and no answers with more than 42 bits are returned. This means that the constraint of never setting multiple bits per clause is fully respected in the evaluation of our QUBO matrix. It is important to note that although there are 5,283 correct solutions in total, these are only distributed across 24 of the 100 randomly generated problem instances. Thus,  most of them have not been solved at all.

\begin{figure}[t]
\centering
\includegraphics[width=.65\textwidth]{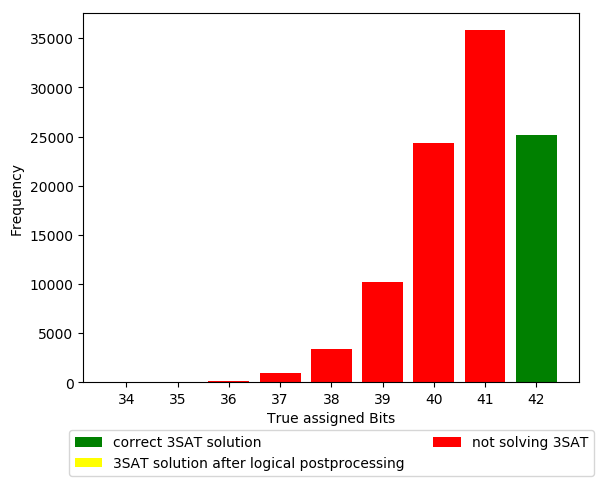}
\caption{Distribution of correct (green) and incorrect (red) answers returned by the quantum annealer \emph{using D-WAVE postprocessing}. Answers that can trivially be transformed into valid answers using logical postprocessing are marked in yellow. The plot shows 100,000 answers in total for 100 different hard 3SAT instances ($\alpha \approx 4.2$).}
\label{fig:distr-yes-pp}
\end{figure}

Furthermore, we applied the logical postprocessing described in Section~\ref{sec:exp-setup} to the incorrect answers in Figure~\ref{fig:distr-no-pp}. However, it shows little improvement on the total amount of correct answers collected. We expect the postprocessing method delivered with the D-Wave software package to be more powerful as it runs local search along more axes of the solution space than the logical postprocessing does. So we ran the complete evaluation experiment again, only this time turning on the integrated postprocessing. The results are shown in Figure~\ref{fig:distr-yes-pp}.

We observed that the D-Wave postprocessing managed to optimize all correct but ``incomplete'' answers, mapping them to a solution with 42 bits assigned the value $\textit{True}$. Out of the 100,000 queries, this yielded 25,142 correct answers. Moreover, these correct answers span 99 of the 100 randomly generated 3SAT instances so that we consider the problem solved. Effectively, this shows that quantum annealing does suffer from a breakdown in expected solution quality at the point of the phase transition in the 3SAT problem. In comparison to the immense decrease in performance seen in classical solvers (cf. Section~\ref{sec:preliminaries}), a drop to around 25\% precision appears rather desirable, though. A quick example: To achieve a $1 - 10^{-12}$ confidence of returning the correct answer our experimental setup requires around $97$ queries. At a glance, that scaling factor with respect to problem difficulty is much better than what is observed for classical algorithms: For example, in the data used for Figure~\ref{fig:crit_sat} we observed performance decrease up to one order of magnitude larger. It is important to note, however, that these experiments were performed for problem instances so small that their evaluation does not pose a challenge to classical processors at all, i.e., below the point of reasonable performance metrics. Thus, these results only proof relevant to practical applications if they scale with future versions of quantum annealing hardware that can tackle much larger problem instances.

\begin{figure}
\centering
\includegraphics[width=.65\textwidth]{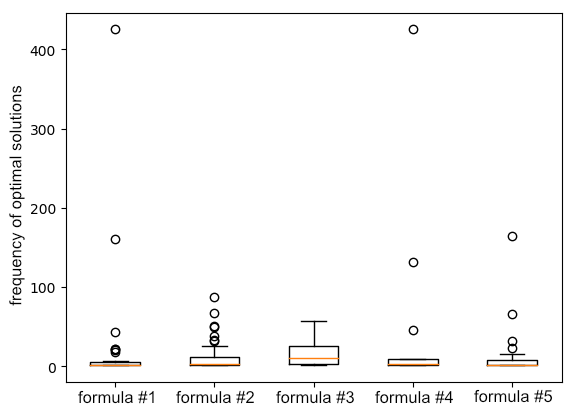}
\caption{Frequency of occurrence of different solutions for 5 formulae with many returned solutions. While most solutions are found once or just a few times, there are specific solutions that are found much more often.} \label{fig:distr-sols}
\end{figure}

So far, we have not discerned between different correct solutions. We were content as long as the algorithm returned but one. However, for the user it is interesting to know if he or she will receive the same solution with every answer or an even distribution across the complete solution space. Our experiments show that when a lot of correct solutions are found for a certain problem instance, there are cases where we can see a clear bias towards a specific solution variant. Figure~\ref{fig:distr-sols} shows the distributions of specific solutions. While some formulae seem to yield rather narrow distributions over the different possible answers, others definitely seem to have a bias towards certain solutions. However, the former also tend to have relatively smaller sample sizes as there are less solutions in total to consider. Further investigation could still reveal a distinctive distribution in these cases as well. Thus, we consider this behavior of the quantum annealer to be roughly in line with the findings of \cite{mandra2017exponentially}, who show an exponential bias in ground-state sampling of a quantum annealer.

\section{Conclusion}
\label{sec:conclusion}


We have shown that problem difficulty of 3SAT instances also affects the performance of quantum annealing as it does for classical algorithms. However, bound by the nature of both approaches, the effects are quite different with complete classical algorithms showing longer runtimes and quantum annealing showing less precision. A first quantification of that loss of precision suggests that it may not be too detrimental and comparatively easy to deal with. However, because of to the maximum available chip size for quantum annealing hardware at the moment, no large-scale test could be performed. No real assumptions on the scaling of this phenomenon (and thus the eventual real-world benefit) can be made yet.

Our results suggest there are cases where single solutions from a set of equally optimal solutions are much more likely to be returned than others. This observation is in line with other literature on the results of quantum annealing. However, it is interesting to note that it translates into the original problem space of 3SAT.

The observed results will gain more practical relevance with larger chip sizes for quantum annealers. We thus suggest to perform these and/or similar tests for future editions of quantum annealing hardware. If the effects persist, they can indicate a substantial advantage of quantum hardware over other known approaches for solving NP-complete problems.

\section*{Acknowledgement}
Research was funded by Volkswagen Group, department Group IT. 
%
%
%
\bibliographystyle{splncs04}
\bibliography{references}

\end{document}